\documentclass[twocolumn,prd,nofootinbib]{revtex4}

\usepackage{graphicx}
\usepackage{amsmath}
\usepackage{amsfonts}
	
\usepackage{epstopdf}

\begin{document}

%............................ definitions ..............
\newcommand{\be}{\begin{equation}}
\newcommand{\ee}{\end{equation}}
\def\bq{\begin{eqnarray}}
\def\eq{\end{eqnarray}}
%........................................................

\title{\bf A Curious Explanation of Some Cosmological Phenomena}
\author{Ram Gopal Vishwakarma}

 \affiliation{Unidad Acad$\acute{e}$mica de Matem$\acute{a}$ticas, Universidad Aut$\acute{o}$noma de Zacatecas,
 C.P. 98068, Zacatecas, ZAC, Mexico, Email: vishwa@matematicas.reduaz.mx}

\begin{abstract}
While observational cosmology has shown tremendous growth over the last decade, deep mysteries continue to haunt our theoretical
understanding of the ingredients of the concordance cosmological model, which are mainly `dark'.
More than 95 percent of the
content of the energy-stress tensor has to be in the form of inflaton, dark matter and dark
energy, which do not have any non-gravitational or laboratory evidence and remain unidentified. Moreover, the dark
energy poses a serious confrontation between fundamental physics and cosmology.
This makes a strong case to discover alternative theories which do not require the dark sectors of the standard approach to explain the observations. 

In the present situation, it would be important to gain insight about the requirements of the `would be' final theory from all possible means.
In this context, the present paper highlights some, hitherto unnoticed, interesting coincidences which may prove useful to develop insight about the `holy grail' of gravitation. 
It appears that the requirement of the speculative dark sectors by the energy-stress tensor, is indicative of a possible way out of the present crisis appearing in the standard cosmology, in terms of a theory wherein the energy-stress tensor does not play a direct role in the dynamics. It is shown that various cosmological observations can be explained satisfactorily in the framework of one such theory $-$ the Milne model, without requiring the dark sectors of the standard approach. Moreover, the model evades  the horizon, flatness and the cosmological constant problems afflicting the standard cosmology.

Though Milne's theory is an incomplete, phenomenological theory, and cannot be the final theory of gravitation, nevertheless, it would be worthwhile to study these coincidences, which may help us develop insight about the would-be final theory.

\noindent
{\bf Key words:}  Theoretical cosmology- Cosmological observations - Alternative explanations.

%\pacs{98.80.-k, 04.20.Cv, 95.30.Sf, 98.80.Jk, 04.20.-q}

\end{abstract}

\maketitle

\section{Introduction}

Theory of general relativity (GR) has been scrutinized by experts for almost a century and is
believed to describe accurately all gravitational phenomena ranging from the solar system to the
universe. However, this success is achieved provided one admits three completely independent
new components in the energy-stress tensor  $-$  infaton, dark matter and dark energy, which
are believed to play major roles in the dynamics of the universe during their turns.  However, there is, until now, no non-gravitational or laboratory
evidence for any of these dark sectors.  Additionally, the mysterious dark energy  poses a serious
confrontation between fundamental physics and cosmology.

Despite the remarkable success of GR, the requirement of the dark sectors is considered by many researchers as a failure of the theory. 
This view is also supported by the apparent failure to unite GR with the quantum field theory describing the other fundamental interactions. Despite the dedicated efforts of more than eighty years, there is still no consensus on how to solve 
and apply quantum gravity principles. 

Amongst the many approaches to quantum gravity proper, one may mention the manifestly covariant
approach of B. S. DeWitt \cite{deWitt}, the use of path integral in Euclidean space advocated by Gibbons
and Hawking \cite{Gibbons&Hawking}, the canonical quantization method of Arnowitt, Deser and Misner \cite{ADM}, the twistor formalism of Penrose \cite{Penrose}, the application of the  path integral technique advocated by Narlikar to quantize the conformal part of the spacetime metric \cite{Narlikar},
the multi-dimensional approach of string theories to reconcile quantum mechanics and GR, giving birth to the 11-dimensional M-theory \cite{Hawking} and finally the attempts to describe the quantum properties of gravity in the loop quantum gravity \cite{LQG}.
Although these methods have emphasized the formal
problems of quantization and led to many interesting abstract concepts, they cannot claim to have
delivered a complete and workable theory of quantum gravity.

 Hence, GR is not and cannot be the final theory of gravitation even if it successfully addresses a wide range of phenomena.
In the present circumstances, it would be important to gain insight about the `would be' final theory from all possible means. Observations can play a crucial role in this direction. Though, the observations are important in all branches of science, they are more important
in a theory of the universe where, on one hand, the events are non-repeatable and, on the other hand, the theoretical
side is more speculative than the laboratory physics, requiring guidelines from the observations
by confronting them. Obviously, one would expect the final theory to explain the observations without requiring the dark sectors of the standard cosmology. 

The present paper unearths some surprising coincidences, observational and theoretical, in the framework of Milne's model, which may prove useful to gain insight about the alternative explanations of the observations without invoking the dark sectors of the standard approach. It would be interesting to note that various cosmological observations can be explained  in the framework of the Milne model, without requiring the dark sectors. Additionally, some long-standing problems of the standard cosmology can also be circumvented in this model. Although, Milne's model does not supply a complete theory of gravitation and is unable to answer why the matter should not curve the spacetime, nevertheless, it would be worthwhile to study the above-mentioned coincidences, which may provide useful clues about the hitherto unknown character of the would-be final theory. As Milne's model is not a widely known theory, we describe briefly its main features in the following.

\section{Milne's Model}

The Milne model  is a  special relativistic cosmological model which was introduced by Edward Arthur Milne in 1935 \cite{milne}. It is a deductive theory based on Milne's kinematic relativity \cite{milneKR} in which information is deduced only from the cosmological principle (together with the basic properties of spacetime and the propagation of light).
The greatest achievement of the kinematic relativity is the possibility of the existence of different time scales. Although there is an infinity of these possible time scales, two are of outstanding importance. One is the local time scale, say $\tau$, 
in which the observers appear to be at rest and the universe presents a static appearance. The second time-scale is universal or cosmic, say $t$, in which the relative motion of the observes is non-zero but unaccelerated (as it is a special relativistic theory). The cosmic time can be identified with the time given by the Robertson-Walker (RW) line element  
\be
ds^2=c^2 dt^2-S^2(t)\left[\frac{dr^2}{1-k r^2}+r^2(d\theta^2+\sin^2\theta ~d\phi^2)\right].\label{eq:RW}
\ee
which is deduced from the assumptions of homogeneity and isotropy as required by the cosmological principle. In order to make the motion of the observers uniform, Milne considered the scale factor $S=ct$ in (\ref{eq:RW}). Now, $k=-1$ is the only choice to make the line element (\ref{eq:RW}) compatible with the Minkowskian metric, since with $S=ct$, the resulting 4-dimensional spacetime from (\ref{eq:RW}) is flat only when $k=-1$ and the 3-space is hyperbolic. Hence, the $t$-time in Milne's model is given by
\be
ds^2=c^2 dt^2-c^2t^2\left[\frac{dr^2}{1+r^2}+r^2(d\theta^2+\sin^2\theta ~d\phi^2)\right].\label{eq:milne}
\end{equation}
One may check that the transformations $\bar{t}=t\sqrt{1+r^2}$, $\bar{r}=ctr$ indeed reduce 
the line element (\ref{eq:milne}) to a manifestly Minkowskian form in the coordinates $\bar{t},\bar{r},\theta,\phi$ (see page 140 in \cite{narlikar}). The $\tau$-time is related with the $t$-time through the transformation
\be
\tau=t_0 \ln\left(\frac{t}{t_0} \right),\label{eq:trans}
\ee
which transforms the line element (\ref{eq:milne}) to a form conformal to a static form of (\ref{eq:milne}):
\be
ds^2=e^{2\tau/t_0}\left[c^2 d\tau^2-c^2t_0^2\left\{\frac{dr^2}{1+r^2}+r^2(d\theta^2+\sin^2\theta ~d\phi^2)\right\}\right],\label{eq:milnen}
\ee
where $t_0$ is a constant with the significance that $\tau=0$ when $t=t_0$. While the line element (\ref{eq:milne}) uses the comoving coordinates and a cosmic time, the metric (\ref{eq:milnen}) uses the locally defined measures of space and time \cite{Bondi}.

Besides the cosmological principle, Milne made another assumption that matter is conserved (which is evidently suggested by ordinary physics). This implies that the equation of hydrodynamic continuity applies and the density of matter decreases with time in the universe whose invariant border advances at the speed of light. 
The zero of $t$-time scale is a fundamental event in the theory when the separation of the fundamental (co-moving) observers vanishes,  proposing a physical explosion of matter. In $\tau$-time scale, this event takes place in the infinite past, owing to its logarithmic dependence on $t$, as is indicated by (\ref{eq:trans}).

It should be noted that the line element (\ref{eq:milne}) results as a natural consequence of kinematic relativity, and has nothing to do with GR. However, as the same solution (\ref{eq:milne}) is obtained in the framework of the standard cosmology for an empty universe, 
it is generally (mis)believed that the Milne model represents an empty universe, which is not correct. All one can say, in the language of GR, is that matter does not curve the spacetime in the geometric analogue of Milne model.
The presence of matter without curving spacetime in Milne's theory, indicates that this theory is fundamentally different from GR and should not be viewed within the usual understanding of an empty universe in GR\footnote{The appearance of a flat spacetime in 
the presence of matter is also not impossible in the conventional GR approach. 
For example, it has been shown in \cite{Ayon-Beato} that 
conformally coupled matter does not always curve spacetime.}. Despite its remarkable success on the kinematic front, the historical development of Milne's model has left the theory in a curiously unfinished state \cite{Bondi}.
When the theory was invented, it met with great hostility and was criticized severely, though often unjustly, and sometimes frivolously.

\section{Compatibility of Milne's Model with Observations}

The standard interpretation of the cosmological observations is provided in the framework of an evolving universe. For this reason, and also to compare the results of the cosmological tests performed on Milne's model vis-a-vis those on the standard cosmology, we consider the $t$-time scale of the Milne model given by the line element (\ref{eq:milne}), in terms of which the universe appears dynamic.
In order to study the cosmological observations, we need to define the luminosity- and the angular diameter- distances in the Milne model. We note that solution (\ref{eq:milne}) provides uniquely, without requiring any input from the matter fields, the luminosity distance $d_{\rm L}$ of a source of redshift $z$ as
\be
d_{\rm L}=cH_0^{-1}(1+z)\sinh\{\ln(1+z)\},\label{eq:d_L}
\ee
where $H_0$ represents  the present value of the Hubble parameter $H=\dot{S}/S$. Hence, the 
angular diameter distance is given by $d_{\rm A}=d_{\rm L}/(1+z)^2$.

\subsection{Supernovae Observations}

Let us first consider the observations of supernovae of type Ia (SNeIa), which render the dark energy as an indispensable ingredient of the standard cosmology. 
An SNIa occurs when a carbon-oxygen white dwarf star in a binary system accretes enough mass from its companion to reach a critical mass and hence undergoes a thermonuclear explosion in its core. Because of the near uniformity of the mass of the white dwarf stars, controlled by the Chandrasekhar limit, the SNeIa produce nearly the same peak luminosity. This allows them to be used as standard candles to measure the distance to their host galaxies because the apparent magnitude of the SNeIa depends primarily on the (luminosity) distance. As the distances are model-based quantities in cosmology and since different cosmological models generally deviate from one another at high redshifts, one can use the high-redshift observations to test and compare the models. As SNeIa are very bright events which can be observed from large cosmological distances (high redshifts), they provide the perfect data for this purpose.

 It is already known that the Milne model, albeit non-accelerating (neither deceleration), is consistent with the observations of SNeIa {\it without requiring any dark energy}.
As early as in 1999, the Supernova Cosmology Project team noticed from the analysis of their first-generation of the SNeIa data that the performance of the empty model ($\Omega_{\rm m}=0=\Omega_\Lambda$) is practically identical to that of the best-fit unconstrained cosmology with a positive $\Lambda$ \cite{perlmutter}.
 Let us consider a newer dataset, for example, the  `new gold sample' of 182 SNeIa \cite{riess}\footnote{Although various newer SNeIa datasets are available, however, the way they are analyzed has left little scope for testing a theoretical model against them. This issue has been addressed by Vishwakarma and Narlikar in \cite{critique}.}, which is a reliable set of SNeIa with reduced calibration errors arising from the systematics. 
It can be checked that the model (\ref{eq:d_L}) provides an excellent fit to the data with a value of $\chi^2$ per degrees of freedom (DoF) $=174.29/181=0.96$ and a probability of goodness of fit $Q=63\%$. Obviously the $\Lambda$CDM model has even a better fit as it has more free parameters: $\chi^2$/DoF $  =158.75/180=0.88$ and $Q=87\%$ obtained for the values $\Omega_{\rm m}=1-\Omega_\Lambda=0.34\pm0.04$. The best-fitting models, the one given by (\ref{eq:d_L}) and the $\Lambda$CDM one, have been compared with this sample of data in Fig. 1.

\begin{figure}
\resizebox{8.5cm}{!}{\includegraphics{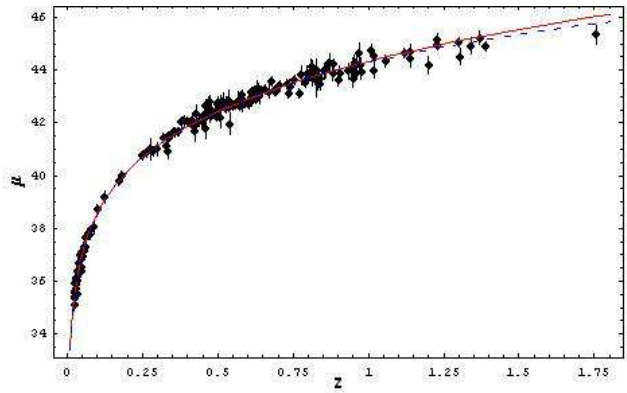}}
{\caption{\small The `new gold sample' of 182 SNeIa from Riess et al. \cite{riess} is compared with some best-fitting models.
The solid curve corresponds to the Milne model and the dashed curve 
corresponds to the spatially-flat $\Lambda$CDM model $\Omega_{\rm m}=1-\Omega_\Lambda=0.34\pm0.04$. 
}}
\end{figure}

\subsection{Observations of High-Redshift Radio Sources}

Let us now consider the data on the angular size and redshift of 256 radio sources with their redshifts in the range 0.5$-$3.8 compiled by Jackson and Dodgson \cite{radio}, which were selected from a bigger sample of 337 ultra-compact radio sources originally compiled by Gurvits \cite{gurvits}. These sources, of angular sizes of the order of a few milliarcseconds
(ultra-compact), were measured by the very long-baseline interferometry. The objects of
the sample of Jackson and Dodgson are short-lived quasars deeply embedded inside the
galactic nuclei, which are expected to be free from evolution on a cosmological time scale
and thus comprise a set of standard rods (at least in a statistical sense).
These sources are distributed into 16 redshift bins, each
bin containing 16 sources. This compilation has recently been used by many authors to test
different cosmological models \cite{radios-used}.

In order to fit this data to the Milne model, let us derive the $\Theta$$-$$z$ relation in the following. The
(apparent) angular size $\Theta$ of a source, of the proper diameter $d$, is given by
\begin{equation}
\Theta(z)=\frac{0.0688dh}{H_0 d_{\rm A}} {\rm milliarcseconds},
\end{equation}
where $d$ is measured in pc, $h$ is the present value of the Hubble parameter in units of 100 km s$^{-1}$ 
Mpc$^{-1}$, and $d_{\rm A}$ is the angular diameter distance given by $d_{\rm A}=d_{\rm L}/(1+z)^2$, as mentioned earlier.

We find that the Milne model has a satisfactory fit to the data with $\chi^2$/DoF $= 20.78/15=1.39$ and $Q = 14\%$.
In order to compare, we find that the best-fitting $\Lambda$CDM model has a slightly better fit: $\chi^2$/DoF $= 16.03/14=1.15$ and $Q = 31\%$ obtained for the values $\Omega_{\rm m}=1-\Omega_\Lambda=0.21\pm0.08$. These models are shown in Fig. 2.

\begin{figure}
\resizebox{8.5cm}{!}{\includegraphics{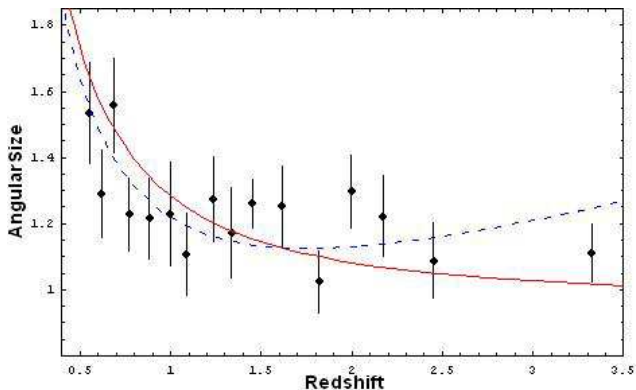}}
{\caption{\small The data on the ultra-compact radio sources compiled by Jackson and Dodgson \cite{radio} is compared with some best-fitting models.
The solid curve corresponds to the Milne model and the dashed curve 
corresponds to the spatially-flat $\Lambda$CDM model $\Omega_{\rm m}=1-\Omega_\Lambda=0.21\pm0.08$. 
}}
\end{figure}

\subsection{Observations of $H_0$ and $t_0$}

The age of the universe $t_0$, in the big bang-like theories,  is the time elapsed since the big bang. It depends on the expansion dynamics of the model and is given by
\begin{equation}
t_0=\int_0^\infty \frac{dz}{(1+z)H(z)}.\label{eq:age}
\end{equation}
Hence, the Hubble parameter controls the age of the universe, which in tern depends on the free parameters of the model.
For example, by the use of the Friedman equation, equation (\ref{eq:age}) reduces to the following in the standard cosmology:
\begin{equation}
t_0=\frac{1}{H_0}\int_0^\infty\frac{(1+z)^{-1}dz}{\sqrt{\Omega_{\rm m}(1+z)^3+\Omega_\Lambda+(1-\Omega_{\rm m}-\Omega_\Lambda)(1+z)^2}}.\label{eq:ageLCDM}
\end{equation}
Although  $t_0$ is a model-based parameter, a lower limit is put on it by requiring that the universe must be at least as old as the oldest object in it. This is done through $t_{\rm GC}$, the age of the globular clusters in the Milky Way which are among the oldest objects we so far know. The parameter $H_0$ can be estimated in a model-independent way, for example, from the observations of the low-redshift SNeIa, in which case the predicted magnitude of the source does not depend on the model-parameters.
 One can use this value to calculate the age of the universe in a particular theory which is to be compared with the age of the oldest objects. Thus the measurements of $H_0$ and $t_{\rm GC}$ provide a powerful tool to test the underlying theory. 

For example, by using the current measurements of $H_0 = 71\pm 6$ km s$^{-1}$ 
Mpc$^{-1}$ from the Hubble Space Telescope Key Project \cite{HST},  equation (\ref{eq:ageLCDM}) gives $t_0$ for the Einstein-deSitter model ($\Omega_{\rm m}=1$, $\Lambda=0$) as $9.18$ Gyr. This cannot be reconciled with the age of the oldest globular cluster estimated to be  $t_{\rm GC}=12.5 \pm 1.2$ Gyr \cite{Gnedin} and the age of the Milky Way as $12.5 \pm 3$ Gyr coming from the latest uranium decay estimates \cite{Cayrel}.
However, for the concordance $\Lambda$CDM model with $\Omega_{\rm m}=1-\Omega_\Lambda=0.27$ (as estimated by the WMAP project \cite{wmap}), equation (\ref{eq:ageLCDM}) gives a satisfactory age of the universe $t_0=13.67$ Gyr which is well above the age of the globular clusters.  Interestingly, the Milne model also qualifies the test: giving a satisfactory age of the universe $t_0=13.77$ Gyr (which is even slightly higher than the concordance model value). 

As has been mentioned earlier, the event $t=0$ in Milne's model, takes place in the infinite past in $\tau$-time scale, owing to a logarithmic dependence of $\tau$ on $t$. Hence, the age of the universe is infinite in $\tau$-time scale.

\subsection{CMB Observations}

Finally, let us see how the Milne model fairs against the observations of the cosmic microwave background (CMB) radiation. The CMB radiation is composed of the photons decoupled from the primordial matter (baryon-photon plasma) at the redshift $z_{\rm dec}$. As the photons were in thermal equilibrium with matter in the plasma before decoupling, the size of the structures in the matter  during the epoch of decoupling (measured by the density contrast), is imprinted on the radiation in the form of small fluctuations in its temperature with respect to a small change in the direction in the sky. Since the universe is optically thin after this epoch, this information remains frozen in the radiation and is seen today. Hence the observed anisotropy in the CMB temperature can be quantified in terms of the size of the structures on the surface of the `last scatter'. For example, a region which has a proper size $L_{\rm dec}$ on  the surface of the last scatter, will subtend an angle $\theta_{\rm dec}$ at the observer today, given by
\be
\theta_{\rm dec}=\frac{L_{\rm dec}}{d_{\rm A}(z_{\rm dec})},\label{eq:theta_dec} 
\ee
where $d_{\rm A}(z_{\rm dec})$ is the (angular diameter) distance to the surface of the last scatter. If one considers $L_{\rm dec}$ to be equal to the Hubble distance $d_{\rm H}(t_{\rm dec})=c H^{-1}(t_{\rm dec})$, equation (\ref{eq:theta_dec}) gives $\theta_{\rm dec}\approx 1^{\rm o}$ for $z_{\rm dec}=1100$ in the standard cosmology. As the CMB appears, in the observations, to be highly isotropic on all angular scales greater than 1$^{\rm o}$, the length scale of the order of $d_{\rm H}(t_{\rm dec})$, is usually interpreted in terms of a horizon (in particular, the sound horizon) giving the largest coherent 
structure in the universe at $t_{\rm dec}$, since this should be the largest distance a sound wave  in the tightly coupled baryon-photon fluid could have traveled since the Big Bang until the epoch of decoupling.

A word of caution is needed here. It should be noted that the Hubble distance defined by $d_{\rm H}(t)=c H^{-1}(t)$ (which is an arbitrary definition) provides only a characteristic distance scale in the universe at $t$ and is {\it not} a horizon, as it does {\it not} have any causal significance (since it does not arise naturally from any light propagation formula). Although in the standard cosmology, the particle horizon (as well as the event horizon) have radii comparable to the Hubble distance $d_{\rm H}$, there are other cosmological models which do not have any horizon. 

Another important point to be noted is that if we interpret  $L_{\rm dec}$, appearing in (\ref{eq:theta_dec}), as the sound horizon, giving the size of the largest coherent region on the last scattering surface in which the homogenizing signals passed at sound speed, then the CMB ought to exhibit large anisotropies ({\it not isotropy}) for angular scales of the order of $1^{\rm o}$ or larger - a result  contrary to what is observed \cite{weinberg, narlikar}.  Hence, it seems that the isotropy of the CMB cannot be explained in terms of some physical process operating under the principle of causality in the standard paradigm \cite{weinberg} (the horizon problem). It is generally believed that inflation made the universe smooth and left the seeds of structures, on the surface of the last scatter, of the order of the Hubble distance at that time\footnote{It may be a matter of debate that if inflation made everything else smooth, why did it leave this significant signature of inhomogeneity (see pages 251, 253 in \cite{narlikar}?}.

Thus all one can say, permitted by the present situation, is that the CMB observations fix a preferred length scale for the size of the structures on the surface of last scatter, which can be estimated in terms of the Hubble distance  $d_{\rm H}(t_{\rm dec})$. For example, this length scale $L_{\rm dec}$ can be written as
\be
L_{\rm dec}=n d_{\rm H}(t_{\rm dec})=n c H^{-1}(t_{\rm dec}),\label{eq:size_dec} 
\ee
where the parameter $n$ can be estimated from the observations of the CMB. 
Particularly, this size can be estimated accurately by using the angular scale of the first peak in the observed angular power spectrum of the CMB, which is supposed to give, with a high precision, the physical scale of the density contrast during the epoch of decoupling. In terms of the  Legendre multipole $\ell$, where $\ell=\pi/\theta$, the WMAP observations \cite{wmap} give the location of the first peak at $\ell=220$. This is equivalent to $\theta_{\rm dec}=0.82^{\rm o}$. Hence, we have to  solve equation (\ref{eq:theta_dec}), taken together with (\ref{eq:size_dec}), for $\theta_{\rm dec}=0.82^{\rm o}$, which would be equivalent to fitting equation (\ref{eq:theta_dec}) (taken together with (\ref{eq:size_dec})) to the first peak in the angular power spectrum of the CMB observed by the WMAP project.

This solution, in the concordance $\Lambda$CDM model ($\Omega_{\rm m}=1-\Omega_\Lambda=0.27$), yields the value $n=0.82$. By considering $H_0 = 71$ km s$^{-1}$ Mpc$^{-1}$, this gives the size of the structures at $t_{\rm dec}$ as $L_{\rm dec}= 182.4$Kpc. For the Milne model, the solution yields $n=7.86$, giving $L_{\rm dec}=30.14$ Mpc. A larger $L_{\rm dec}$, in the Milne model, is a characteristic of a higher expansion rate, which also results in a larger distance to the surface of the last scatter $d_{\rm A}(z_{\rm dec})=2.11$ Gpc, compared to $d_{\rm A}(z_{\rm dec})=12.78$ Mpc in the concordance model. The length scale $L_{\rm dec}$ is expected to grow, due to the cosmic expansion, to a proper length $L_0$ today, given by $L_0=(1+z_{\rm dec})L_{\rm dec}$. Thus the present size of our patch of homogeneity and isotropy, is 33.19 Gpc in the Milne model compared with 200.82 Mpc in the concordance model.

It is known from observations that the present size of our patch of
approximate large-scale homogeneity and isotropy is at least as big as the present-day
Hubble distance\footnote{As the size of the horizons in the standard cosmology is of the order of the Hubble distance, the size of the observable universe is regarded to be of the order of the Hubble distance ($\approx 4.2$ Gpc).}.
It would be worthwhile to mention here that Grishchuk \cite {grishchuk} has found out, by combining available observations with plausible statistical assumptions, that the present size of this patch is significantly bigger than the present Hubble distance. In this context, the value of this patch as calculated in the Milne model - about an order of magnitude higher than that in the standard cosmology - is encouraging.

Thus, there seems a possibility, in the limited scope given by the Milne model, to explain the location of the first peak in the observed angular power spectrum of CMB. Moreover, the size of the observable universe predicted by this model seems consistent with Grishchuk's findings. Nevertheless, it is not possible to explain and quantify the generation of the acoustic oscillations and the locations of the other peaks in the framework of Milne's model, which lacks the early universe physics. Hence, the overall explanation of CMB cannot be considered satisfactory in the Milne model, compared with the standards of the corresponding explanations in the concordance cosmology.

\section{On the Problems of the Standard Cosmology}

Let us now witness some coincidences in the Milne model registered on the theoretical front.  As we see in the following, the model can circumvent the long-standing problems of the standard cosmology, for example, the horizon, flatness and the cosmological constant problems.

\bigskip
\noindent
{\bf Horizon Problem:} 

\medskip
\noindent
The distance of the (particle) horizon, given by
\be
d_{\rm h}(t)=S(t) \int^t_0\frac{cdt'}{S(t')},\label{eq:hor}
\ee
sets a limit of the observable or the causally connected part of the universe at time $t$. As a finite value exists for $d_{\rm h}$ in the standard cosmology, this means that the universe has a horizon in this theory. This is in conflict with the observed smoothness of the CMB at the largest scales in all directions, indicating that even the parts of the universe lying outside the horizon have been in thermal contact. While, the standard cosmology has to take refuge in inflation  in order to solve this problem, the problem does not exist in the Milne model, as $d_{\rm h}=\infty$ at all times in this model (as can be checked from (\ref{eq:milne}) and (\ref{eq:hor})), and the universe is always causally connected.

\bigskip
\noindent
{\bf Cosmological Constant Problem:}

\medskip
\noindent
 In the standard cosmology, the origin of the cosmological constant problem lies in a conflict between the values of the cosmological constant and the energy density of vacuum in the quantum field theory (QFT). The vacuum energy, according to the QFT,
results from the quantum vacuum fluctuations which provide an energy contribution of the
order of the Planck mass. In GR, the vacuum energy can be represented by the cosmological constant.  Friedman equation 
 then provides an estimate of the vacuum energy in terms of $H^2_0$. This is, however, smaller
than the QFT-value by a factor of $\approx 10^{120}$!  This discrepancy has been called `the worst
theoretical prediction in the history of physics!' This problem is evaded in Milne's model owing to the fact that neither the cosmological constant, nor any other dark energy candidate appears in the theory.

\bigskip
\noindent
{\bf Flatness Problem:} 

\medskip
\noindent
The standard cosmology harbours this problem through the Friedmann equation 
\be
\frac{\rho}{3H^2/(8\pi G)} -1\equiv\frac{\rho}{\rho_{\rm c}}-1\equiv\Omega-1=\frac{kc^2}{S^2H^2},\label{eq:friedmann}
\ee
implying that the universe will have positive, zero or negative spatial curvature depending on whether its total energy density $\rho$ is more than, equal to or less than the critical density $\rho_{\rm c}$. 
As $|\Omega-1|$ grows with time\footnote{For instance, in the standard cosmology with $S\propto t^{1/2}$, one has $|\Omega-1|\propto t$.} according to (\ref{eq:friedmann}), this causes a problem that even a minute departure of early $\Omega$ from unity grows significantly in time  and yet the universe today remains very close to flat.
For example, the observational uncertainty of $\Omega$ at present, would require it to be differing from unity by $10^{-53}$ during the GUT epoch! Any relaxation of this fine tunning would have led to a far wider range of $\Omega$ at present than is permitted by the observations. 

Equation (\ref{eq:friedmann}), for dust, can also be derived in the kinematic theory by using the continuity equation and the Navier-Stokes equation of fluid dynamics (see, for example, pages 125-127 of \cite{narlikar}), though with a different meaning of its terms:
\be
\frac{\dot{S}^2}{2}+\frac{-G\frac{4\pi S_0^3}{3}\rho_0}{S}=-\frac{kc^2}{2}.\label{eq:energies}
\ee
The two terms on the left are respectively the kinetic energy and the gravitational potential energy (per unit mass) in the universe. Hence, unlike GR, here the constant $k$ is related with the total energy of the universe, and not with its curvature. We can relate it to the total energy of the universe at the present epoch. 

The situation differs here from the standard cosmology case in the fact that the right hand side of (\ref{eq:energies})) does not evolve with time, unlike the case in (\ref{eq:friedmann}). Hence, any fine tunning between the two energy terms is not required in Milne's model,  and their sum remains constant. Moreover, the most likely value of the constant $k$ in (\ref{eq:energies})) is zero, as many theoretical findings claim that the total energy of the universe should be zero.

\medskip
\section{Conclusion}

It appears that the dark ingredients of the energy-stress tensor - the inflaton field, the non-baryonic dark matter and the dark energy - have become more like liabilities than assets of the standard cosmology.
One may surmise that the requirement of the speculative dark sectors by the energy-stress tensor, is indicative of a possible way out of the present crisis (in the standard cosmology), in terms of a theory wherein the energy-stress tensor is absent.

 It would be interesting to note that various cosmological observations can be explained successfully in the framework of one such theory - the Milne model.
 Additionally, the Milne model evades  the horizon, flatness and the cosmological constant problems afflicting the standard cosmology.
However,  an alternative theory cannot be acceptable purely on the basis of its success on the largest scales. It is also expected to pass the tests through the local observations, for example those which have been devised to test GR. Clearly, Milne's theory appears far from meeting this challenge. 

Nevertheless, the various theoretical and observational coincidences studied in the framework of Milne's model are worth paying attention to, owing to the fact that they do not require any of the dark sectors of the standard cosmology.
This may contain some, hitherto unnoticed, important clues about the final theory of gravitation.

\bigskip
\noindent
{\bf Acknowledgement:} The author thanks an anonymous referee for making some constructive comments which though made me work overtime, but improved the quality of the paper significantly.


\newpage

\bigskip
\noindent
\begin{thebibliography}{X}



\bibitem{deWitt} B.S. DeWitt, Phys. Rev. 162 (1967) 1239.

\bibitem{Gibbons&Hawking} S.W. Hawking, in {\it General Relativity - An Einstein Centenary Survey}, eds. SW. Hawking and W. Israel (Cambridge, 1979).

\bibitem{ADM} R. Arnowitt, S. Deser and C.W. Misner, in {\it Gravitation - An introduction to current research}, ed. L. Witten (John Wiley, New York, 1962).

\bibitem{Penrose} R. Penrose, in {\it Quantum Gravity - An Oxford Symposium}, eds. C.J. Isham, R. Penrose and D.W. Sciama (Oxford University Press, 1975).

\bibitem{Narlikar} J. V. Narlikar, Foundations of Physics, Vol. 14, 443, 1984.

\bibitem{Hawking} S.W. Hawking and L. Mlodinow, {\it The Grand Design}, (Bantam Books, New York, 2010).

\bibitem{LQG} C. Rovelli,  {\it Loop Quantum Gravity}, Physics World, November 2003. 

\bibitem{milne} E. A. Milne, {\it Relativity, Gravitation and World Structure}, (Oxford, 1935).

\bibitem{milneKR} E. A. Milne, {\it Kinematic Relativity}, (Oxford, 1948).

%\bibitem{milne&walker} E. A. Milne and A. G. Walker,  Observatory, {\bf 64}, 17, 1940.

\bibitem{narlikar} J. V. Narlikar, {\it An Introduction to Cosmology} (Cambridge University Press, 2002).

\bibitem{Ayon-Beato} E. Ayon-Beato, C. Martinez, R. Tronoso and J. Zanelli, Phys. Rev. D, {\bf 71}, 104037, 2005.

\bibitem{Bondi} H. Bondi,  {\it Cosmology} (Cambridge University Press, second edition, 1968).

\bibitem{perlmutter} S. Perlmutter, et al., Astrophys. J. {\bf 517}, 565,1999.

\bibitem{riess} A. Riess, et al., Astrophys. J. {\bf 659}, 98, 2007.

\bibitem{critique} R. G. Vishwakarma and J. V. Narlikar, Res. Astron. Astrophys. {\bf 10}, 1195, 2010. 

\bibitem{radio} J. C. Jackson and M. Dodgson, Mon. Not. R. Astron. Soc. {\bf 285}, 806, 1997.

\bibitem{gurvits} L. I. Gurvits, Astrophys. J. {\bf 425}, 442, 1994.

\bibitem{radios-used} S. K. Banerjee and J. V. Narlikar, Mon. Not. R. Astron. Soc. {\bf 307}, 73, 1999;\\
R. G. Vishwakarma, Class. Quantum Grav. {\bf 17} 3833, 2000;\\
R. G. Vishwakarma and Parampreet Singh, Class. Quantum Grav. {\bf 20}, 2033, 2003;\\
R. G. Vishwakarma, Nuovo Cim. B {\bf 122}, 113, 2007. 

\bibitem{HST} J. R. Mould, et al., Astrophys. J. {\bf 529}, 786, 2000.

\bibitem{Gnedin} O. Y. Gnedin, O. Lahav and M. J. Rees, preprint: astro-ph/0108034.

\bibitem{Cayrel} R. Cayrel, et al., Nat. {\bf 409}, 691, 2001.

\bibitem{wmap} D. Larson,et al., Astrophys. J. Suppl. {\bf 192}, 16, 2011.

\bibitem{weinberg} S. Weinberg, {\it Gravitation and Cosmology: Principles and Applications of the General Theory of Relativity} (John Wiley \& Sons, 1972), p 525.

%\bibitem{narlikar_cmb} J. V. Narlikar,  {\it An Introduction to Cosmology} (Cambridge University Press, 2002), p 251, 253.

%\bibitem{grishchuk_cmb} L. P. Grishchuk, {\it Space Sci. Rev.}, {\bf148}, 315, 2009.

\bibitem{grishchuk} L. P. Grishchuk, {\it Phys. Rev. D} {\bf 45}, 4717, 1992.

%\bibitem{vishwa} R. G. Vishwakarma, Astrophys. Space Sci. {\bf 340}, 373, 2012.


\end{thebibliography}
\end{document}